\documentclass{article}

\usepackage{arxiv}

\usepackage[utf8]{inputenc} 
\usepackage[T1]{fontenc}    
\usepackage{hyperref}       
\usepackage{url}   
\usepackage{amsmath}
\usepackage{booktabs}       
\usepackage{amsfonts}       
\usepackage{nicefrac}  
\usepackage{subfig}
\usepackage{microtype}      
\usepackage{lipsum}		
\usepackage{graphicx}
\usepackage{natbib}
\usepackage{doi}

\newcommand{\s}{\mathbf{s}} 
\newcommand{\coh}{\mathbf{s}} 
\newcommand{\p}{\mathbf{p}} 
\newcommand{\x}{\mathbf{x}} 
\newcommand{\w}{\mathbf{w}} 
\newcommand{\y}{\mathbf{y}} 
\newcommand{\X}{\mathbf{X}} 
\newcommand{\W}{\mathbf{W}} 
\newcommand{\z}{\mathbf{z}} 
\newcommand{\m}{\mathbf{m}} 
\newcommand{\mc}{{\boldsymbol\mu}} 
\newcommand{\varc}{{\mathbf{V}}} 
\newcommand{\KL}[2]{D_{\mathrm{KL}}\left(#1 \parallel #2\right)} 
\newcommand{\KLk}{H} 

\newcommand{\g}{\mathbf{g}} 
\newcommand{\f}{\mathbf{f}} 

\newcommand{\N}{\mathcal{N}} 

\title{On extracting coherent seismic wavefield using variational symmetric autoencoders}

\author{ \href{http://orcid.org/0000-0003-4081-8969}{\includegraphics[scale=0.06]{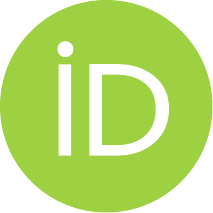}\hspace{1mm}Pawan Bharadwaj}\\
	Centre for Earth Sciences\\
	Indian Institute of Science\\
	Bengaluru \\
	\texttt{pawan@iisc.ac.in} \\
}

\renewcommand{\shorttitle}{Variational Symmetric Autoencoders}

\hypersetup{
pdftitle={On extracting coherent seismic wavefield using variational symmetric autoencoders},
pdfauthor={Pawan Bharadwaj},
pdfkeywords={ coherent wavefield; seismic deconvolution; autoencoders; probabilistic models; deep learning; neural networks; unsupervised learning; representation learning; earthquakes; source estimation},
}

\begin{document}
\maketitle

\begin{abstract}
	
Many seismological applications, such as receiver function analysis, earthquake source imaging, and ambient noise tomography, rely on averaging waveforms to extract coherent information. While traditional linear stacking methods have been widely used, recent studies have demonstrated the advantages of nonlinear stacking techniques in enhancing signal-to-noise ratios. Our work effectively generalizes this concept and
enables a more sophisticated form of non-linear stacking, using probabilistic generative models.
We discuss the variational formulation of the Symmetric Autoencoder (SymAE) and its role in achieving disentanglement within the latent space to extract coherent information from a collection of seismic waveforms. Disentanglement involves separating the latent space into components for coherent information shared by all waveforms and components for waveform-specific nuisance information.
SymAE employs a generative model that independently generates waveforms based on coherent and nuisance components, and an inference model that estimates these components from observed wavefield. By assuming the independence of waveforms conditioned on coherent information, the model effectively accumulates this information across multiple waveforms. 
After training, a metric based on Kullback-Leibler divergence is used to evaluate the informativeness of individual waveforms, enabling latent-space optimization and the generation of synthetic seismograms with enhanced signal-to-noise ratios.

To demonstrate the efficacy of our proposed method, we applied it to a data set of teleseismic displacement waveforms of the P wave from 
deep-focus earthquakes. By training the SymAE model on high-magnitude events, we successfully identified seismograms that contained robust source information.  Furthermore, we generated high-resolution virtual seismograms enriched with relevant coherent source information and less influenced by scattering noise, allowing a deeper understanding of the characteristics of the earthquake source. Importantly, our method extracts coherent source information without relying on deconvolution, which is often used in traditional source imaging. This enables the analysis of complex earthquakes with multiple rupture episodes, a capability that is not easily achievable with conventional approaches.

\end{abstract}


\section{Introduction}
Probabilistic generative models~\citep{goodfellow2020generative, kingma2019introduction, ho2020denoising, rezende2015variational} have emerged as versatile tools across numerous scientific disciplines. Their applications span from basic image processing tasks such as compression, denoising, and inpainting to more complex problems like semi-supervised and unsupervised learning. While the core principles remain consistent, their implementation and application vary significantly across different fields.
Recent advancements have witnessed a growing interest in applying generative models to geophysical data~\citep{song2021bridging, xiao2024diffusion, trappolini2024cold, ren2024learning, sun2024enabling}. 
In geophysics, a crucial aspect of generative model application lies in the physical interpretability of the generated data. In other words, the ability to generate samples conditioned on physically relevant variables is often paramount. This necessitates the development of generative models that can not only capture the statistical properties of the data, but also produce physically meaningful and informative outputs.

This study introduces a novel probabilistic generative model, the Symmetric Autoencoder (SymAE), designed to extract  coherent information from seismological data.
Traditional methods for extracting coherent energy often employ preprocessing steps, such as phase correction and linear stacking, to suppress nuisance variations. Recent advancements have introduced non-linear stacking~\citep{schimmel2011using,olivier2015body,ruckemann2012comparison,weaver2018temporally,xie2020improving} techniques that assign weights to waveforms based on quality assessments. Our approach extends this concept by performing a non-linear decomposition of waveforms into constituent components. Subsequently, a neural network learns optimal weights for each component, enabling a more sophisticated form of non-linear stacking.
We refer to this process as accumulating coherent information across waveforms, which we argue is more effective than traditional stacking methods.

SymAE~\citep{mlvae, bharadwaj2022redatuming} was developed to learn a disentangled representation of the seismic wavefield to generate waveforms conditioned on coherent information relevant to the task. However, generating waveforms enriched in coherent information while minimizing the effects of task-irrelevant nuisances was not addressed. This work introduces a metric for quantifying waveform informativeness based on its contribution to accumulated coherent information, enabling the generation of relevant waveforms. We pose the extraction of coherent information as a latent-space optimization problem. This process is data-driven and does not rely on knowledge of the underlying physical signal model.
Additionally, an upgrade to SymAE that is based on spatial transformer networks~\citep{jaderberg2015spatial} and attend-infer-repeat networks~\citep{eslami2016attend} is proposed to handle traveltime shifts in the observed time domain waveforms, potentially obviating the need for preprocessing steps like cross-correlation.

\begin{figure}
  \centering
  \includegraphics[width=\linewidth]{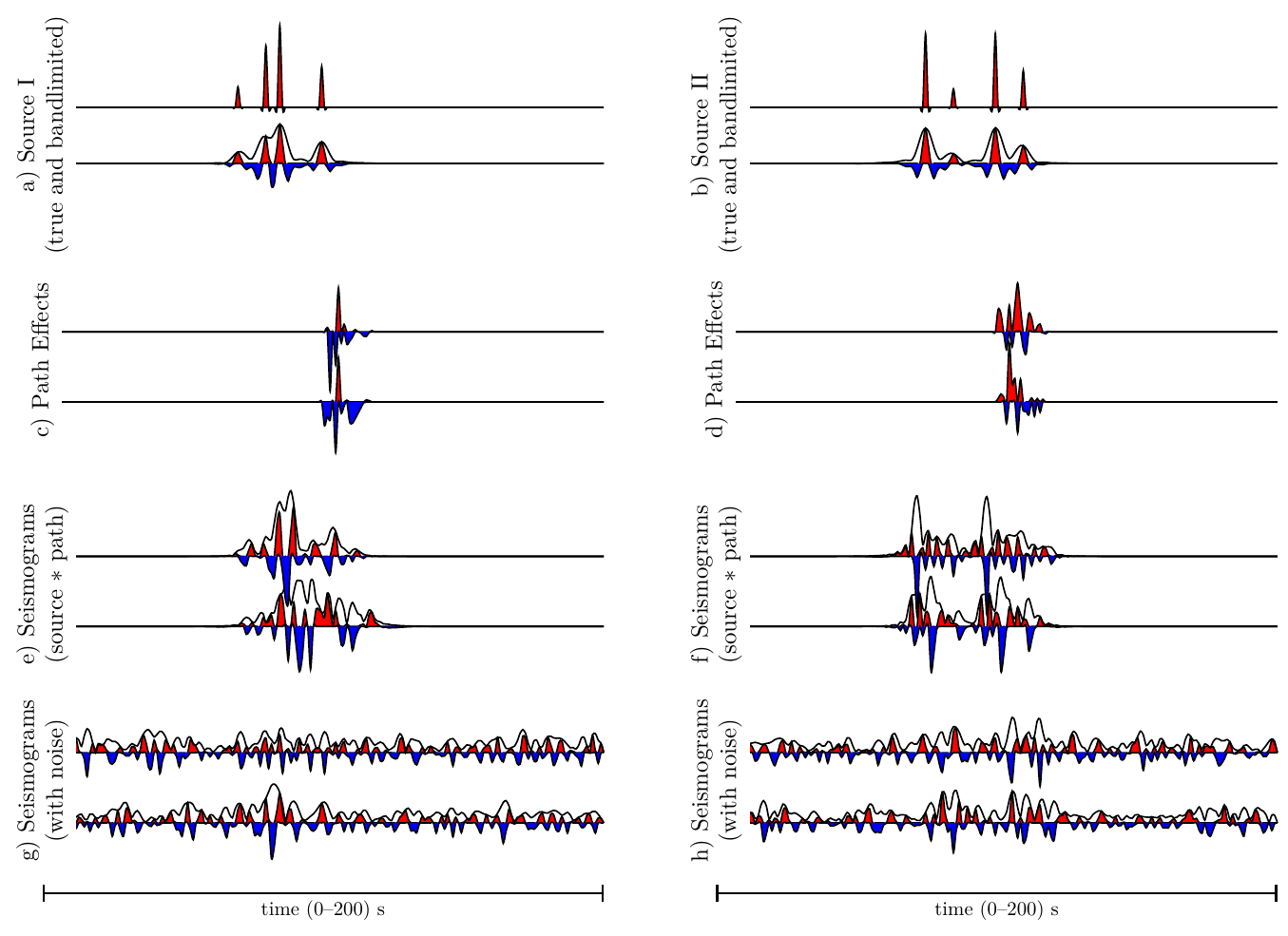}
  \caption{
  Extracting coherent source information from seismograms modeled as the convolution of band-limited source signatures with path effects, corrupted by noise and time shifts.
}
  \label{fig:syn_exp1}
\end{figure}

\begin{table}
\centering
\begin{tabular}{|cccccccc|}
\toprule
\hline
\# & Code &       Date &     Time &  Mw &  Depth &     Half Duration & Region \\ \hline
\midrule
     1 & okt1 & 2013-05-24 & 05:44:48 & 8.30 & 598.00 & 35.7 &      Sea of Okhotsk \\
    2 & okt2 & 2013-05-24 & 14:56:31 & 6.70 & 624.00 &  5.6 &           Sea of Okhotsk \\
    3 & okt3 & 2008-07-05 & 02:12:04 & 7.70 & 633.00 & 17.3 &             Sea of Okhotsk \\
    4 & okt4 & 2008-11-24 & 09:02:58 & 7.30 & 492.00 & 10.9 &            Sea of Okhotsk \\
    5 & okt5 & 2013-10-01 & 03:38:21 & 6.70 & 573.00 &  5.6 &             Sea of Okhotsk \\
    6 & okt6 & 2012-08-14 & 02:59:38 & 7.70 & 583.00 & 17.8 &           Sea of Okhotsk \\
    7 &  pb2 & 2015-11-24 & 22:45:38 & 7.60 & 606.00 & 16.6 &                Peru Brazil \\
    8 & fiji1 & 2018-08-19 & 00:19:40 & 8.20 & 600.00 & 30.8 &             Fiji Islands  \\
    9 & fiji2 & 2018-09-06 & 15:49:18 & 7.90 & 670.00 & 21.4 &              Fiji Islands  \\
    10 & fiji3 & 2014-11-01 & 18:57:22 & 7.10 & 434.00 &  -   &              Fiji Islands  \\
    11 & fiji4 & 2018-11-18 & 20:25:46 & 6.80 & 540.00 &  5.8 &               Fiji Islands  \\
    12 & fiji5 & 2021-04-24 & 00:23:38 & 6.50 & 301.00 &  4.5 &           Fiji Islands  \\
    13 & bon1 & 2015-05-30 & 11:23:02 & 7.80 & 664.00 & 20.7 &     Bonin Island Japan Reg \\
    14 & van1 & 2014-03-05 & 09:56:57 & 6.30 & 638.00 & 3.8  &           Vanuatu Islands \\
    15 &  spain & 2010-04-11 & 22:08:12 & 6.30 & 609.00 &  -   &                     Spain \\
    16 &  nbl & 1994-06-09 & 00:33:16 & 8.20 & 631.00 & 20.0 &            Northern Bolivia \\
    \hline
\bottomrule
\end{tabular}
\caption{\label{tab:eq_details}
SymAE was trained using displacement seismograms of these deep focus earthquakes.
}
\end{table}
Our numerical experiments concentrate on the unsupervised extraction of coherent earthquake source information~\citep{dziewonski1981determination, sipkin1982estimation, abercrombie2015investigating} from a collection of seismograms. The seismograms of deep focus earthquakes, listed in Tab.~\ref{tab:eq_details}, used are obscured by path scattering effects. After learning a representation of seismograms through SymAE, we demonstrate its application to quantify scattering effects in individual seismograms. Then, we generate seismograms with reduced path-related noises to gain deeper insight into the characteristics of the earthquake sources.
In addition to real earthquake data, we considered two synthetic earthquake sources for generating training and testing datasets with roughly 5000 seismograms. These synthetic sources, along with representative path effects and sample seismograms, are illustrated in Fig.~\ref{fig:syn_exp1}. To generate the seismograms, we convolved the synthetic sources with path effects representing P-arrivals and their coda. Random time shifts, band-limited noise, and normalization were applied to simulate realistic traveltime errors and noises commonly observed in seismological data. 
The purpose of the synthetic experiment is to retrieve coherent source time functions from noisy seismograms, as plotted in Figs.~\ref{fig:syn_exp1}g and \ref{fig:syn_exp1}h, without deconvolution using known path effects. To the best of our knowledge, this marks the first work to discuss the extraction of earthquake source information in an entirely model-free manner.

While this study focuses on earthquake source estimation, the proposed methodology is applicable to a broader range of seismological applications. Many seismic tasks involve extracting coherent information from waveforms, such as receiver function analysis, earthquake source imaging, and ambient noise tomography. These applications often rely on grouping waveforms based on shared characteristics, either source-related or path-related.
\begin{itemize}

\item Receiver function analysis~\citep{Zandt1995, Zheng2018, Dalai2021, Bloch2023}, where coherent information on the near-receiver Earth's crustal structure should be extracted from waveforms related to numerous teleseismic earthquake events. Compared to linear stacking~\citep{Hu2015}, where numerous events are required, a non-linear method to extract
coherent information can allow for investigation of azimuthal and epicentral-distance dependence (attributed to
anisotropy and 3D orientation) of the receiver functions.

\item 
Coherent information from 
station-pair cross-correlations~\citep{weaver2001lobkis,weaver2005information, campillo2003long,wapenaar2004retrieving,snieder2004extracting} is extracted linearly, where the emergence of this information is proportional to the square root of the recording time and inversely proportional to the square root of the distance between stations. 
The coherent information retrieved in this scenario pertains to the medium between the stations and is utilized for tomography purposes.

\end{itemize}

This work utilizes a $D$-dimensional vector $\x_j^k$
to represent the measured seismic waveform in the time domain. The subscript $j$ indicates the source of the waveform (the $j$th earthquake), and the superscript $k$ indicates the location where it was measured (the $k$th receiver).
To represent a group of waveforms recorded due to a single earthquake source (the $j$th earthquake) and measured at different receivers, we denote it as a set $\X_j=\{\x_j^k\}$, where we simply drop the receiver index $k$ and use a capital letter.
Although this paper focuses mainly on source-based groups $\X_j$, we can similarly establish a notation for receiver-based collections by omitting the source index, denoted as $\X^k$, to tailor the equations presented here to the application scenarios discussed previously.
Finally, to denote a group of waveforms for all sources and receivers, we remove the superscript and use the subscript to access the elements, i.e. $\X=\{\x_i\}$ represents the group of all measured waveforms,
where $i$ iterates over all the source-receiver combinations.

\section{Variational Autoencoder}

Variational autoencoders~\citep[VAE]{kingma2013auto, kingma2019introduction, prince2023understanding, bishop2023deeplearning, higgins2016beta, burgess2018understanding} are a class of probabilistic generative models designed to infer the distribution $P(\X)$ from the observed wavefield.
VAE is described as a nonlinear latent variable model, which expresses the probability distribution of each element $\x_i$ as an integral over the latent variable $\z$ of the joint probability distribution of $\x_i$ and $\z$:
\begin{equation}
P(\x_i) = \int P(\x_i, \z)\,\text{d}\z = \int P(\x_i\mid\z) P(\z)\,\text{d}\z.
\end{equation}
Here, the \emph{likelihood} $P(\x_i\mid\z)$ is the conditional probability of $\x_i$ given $\z$, modeled as a normal distribution $\N$, where its mean and diagonal covariance matrices are produced by a decoder function $\f$ of $\z$ and parameters $\phi$:
$P(\x_i\mid\z,\phi) = \N(\x_i\mid\mathbf{f}[\z,\phi])$. In this work,
we parameterize $\mathbf{f}$ as a neural network with convolutional layers. 
The \emph{prior} probability of $\z$, denoted as $P(\z)$, is also assumed to follow a normal distribution with zero mean $\mathbf{0}$ and the identity covariance matrix $\mathbf{I}$: $P(\z) = \N(\z\mid\mathbf{0}, \mathbf{I})$. 
In seismology, the latent or hidden variables can be considered as either the path or source parameters that must be specified to produce the wavefield. In this sense, the decoder $\f$ can be seen as a network performing the \emph{forward modeling} operation given the latent variables.

In order to determine the decoder parameters $\phi$, which control the generation of the wavefield based on the latent variable setting, VAE aims to maximize the log likelihood of the wavefield $\mathbf{X}$. Assuming independently and identically distributed waveforms, we can write:
\begin{equation}
\label{eqn:latentmodel}
\log[P(\mathbf{X}\mid\phi)] = \sum_i \log[P(\x_i\mid\phi)] = \sum_i \log\left[\int P(\x_i\mid\z,\phi) P(\z) \text{d}\z\right].
\end{equation}
Since the log likelihood expressed as an integral over $\z$ is intractable~\citep{bishop2023deeplearning}, we decompose it into two terms:
\begin{eqnarray}
  \log[P(\mathbf{X}\mid\phi)] =
  \sum_i {\int Q_i(\z) \log\left[\frac{P(\x_i\mid\z,\phi)P(\z)}{Q_i(\z)}\right] \text{d}\z}
  + \sum_i \KL{Q_i(\z)}{P(\z\mid\x_i,\phi)}.
  \label{eqn:llh1}
\end{eqnarray}
It can be shown that this decomposition holds for any auxiliary distribution $Q_i(\z)$, and
it is commonly employed within the framework of variational inference to facilitate the approximation of intractable integrals. 
In eq.~\ref{eqn:llh1},
$P(\z\mid\x_i,\phi)$ is the \emph{posterior}
distribution of $\z$ given $\x_i$ --- it specifies the distribution of the path and the source latent variables responsible for the generation of the 
waveform $\x_i$. 
The second term here gives a summation over the Kullback-Leibler (KL) divergence
\begin{eqnarray}
\KL{Q_i(\z)}{P(\z\mid\x_i,\phi)} = \int Q_i(\z) \log\left[\frac{Q_i(\z)}{P(\z \mid \x_i, \phi)}\right] \mathrm{d}\z,
\end{eqnarray}
which measures the difference between each auxiliary distribution $Q_i(\z)$ and the posterior distribution $P(\z \mid \x_i, \phi)$. As the KL divergence is always non-negative (can be shown using Jensen's inequality), the first term in eq.~\ref{eqn:llh1}, which is the Evidence Lower Bound (ELBO), serves as a lower bound on the log-likelihood of the data. 
VAE focuses on the maximization of this term, given that the estimation of the posterior $P(\z\mid\x_i,\phi)$ in the second term remains intractable.
It has to be noted that the maximization of the ELBO inherently facilitates the minimization of the KL divergence between $Q_i(\z)$ and $P(\z\mid\x_i,\phi)$.
%

VAE utilises an encoder network $\mathbf{g}$, parameterized by $\theta$, to perform amortized inference of $\z$. 
Specifically, the network $\mathbf{g}$ generates auxiliary distributions that are conditioned on the input waveform $\x_i$. These auxiliary or approximate posterior distributions are assumed to be Gaussian distributions. The mean and diagonal covariance of these Gaussian distributions are determined by the output of the network $\mathbf{g}$ when given the input $\x_i$ and the parameters $\theta$. This is mathematically represented as $Q_i(\z) = Q(\z\mid\x_i,\theta) = \N(\z\mid\mathbf{g}[\x_i,\theta])$.
The underlying rationale is that the encoder function $\g$, again parameterized using convolutional layers in our case, performs \emph{inverse modelling} to estimate the distribution of the source and path parameters given a wavefield $\x_i$.
Finally, this choice of $Q_i(\z)$ makes the Evidence Lower Bound (ELBO) in Eq.~\ref{eqn:llh1} for VAEs a function of the parameters \(\phi\) and \(\theta\), which is maximized to determine their optimal values:
\begin{equation}
\label{eqn:ELBO1}
\mathbb{L}(\theta, \phi) = \sum_i \int Q(\z\mid\x_i, \theta) \log\left[\frac{P(\x_i\mid\z,\phi)P(\z)}{Q(\z\mid\x_i, \theta)}\right] \mathrm{d}\z.
\end{equation}
Using the product rule of the logarithm, the 
ELBO during the training process is further decomposed into two terms: 
\begin{equation}
\label{eqn:ELBO2}
\mathbb{L}(\theta, \phi) = \sum_i \int Q(\z\mid\x_i, \theta) \log\left[P(\x_i\mid\z,\phi)\right] \mathrm{d}\z - \sum_i \KL{Q(\z\mid\x_i, \theta)}{P(\z)}.
\end{equation}
The first term 
represents the reconstruction loss, which measures how accurately the model can reconstruct the original data from the latent variables.
During training, the reconstruction loss term is approximated using a Monte Carlo estimate, such that 
\begin{equation}
\label{eqn:ELBO3}
\mathbb{L}(\theta, \phi) = \sum_i\log[P(\x_i\mid\z^{*},\phi)] - \sum_i\KL{Q(\z\mid\x_i, \theta)}{P(\z)},
\end{equation} 
where $\z^{*}$ denotes a sample drawn from $Q(\z\mid\x_i, \theta)$. 
Maximizing this term improves the model's ability to reproduce the original data. 
The second term in eq.~\ref{eqn:ELBO2} acts as a regularizer, minimizing the KL divergence between the approximate posterior distribution $Q(\z\mid\x_i, \theta)$ and the prior distribution $P(\z)$.

In summary, 
VAE is a type of generative model that is well suited to learning latent representations of seismic data. They consist of two key components: an encoder $\g$ and a decoder $\f$. The encoder maps the input waveforms to a compressed latent-space representation $\z$.
Although VAEs are capable of producing virtual waveforms by drawing samples from the aggregated posterior distribution $\frac{1}{I}\sum_{i=1}^{I}Q(\z\mid\x_i, \theta)$ of $I$ waveforms, and then processing these samples through the decoder, their utility is limited. The virtual waveforms generated by a basic VAE often lack a clear physical interpretation, making it difficult to extract task-relevant information.
To address this limitation, recent research has focused on learning disentangled representations within the VAE framework~\cite{burgess2018understanding}. 
The next section will discuss learning such disentangled representations using the symmetric autoencoder.

\section{Variational Symmetric Autoencoder}

This section discusses the variational formulation of the Symmetric Autoencoder (SymAE). We will explore how SymAE achieves disentanglement within the latent space, ultimately enabling the extraction of coherent information from seismic wavefield data.
Disentanglement refers to the process of separating the latent space into distinct factors that correspond to physically meaningful components of the wavefield. This translates to separating the latent space into factors representing coherent source information and nuisance path information. 
We consider the task of representing each group $\X_j$ of seismograms, corresponding to a particular earthquake, using a latent space variable $\z$ partitioned into the following components:
\begin{itemize}
\item Source component ($\s$): this component captures the source information accumulated from all the seismograms of $\X_j$. It is shared by all the seismograms within the group, reflecting the common signal originating from the earthquake itself.
\item Path component ($\p^k$): this component represents nuisance information, specifically the path scattering effects unique to each individual seismogram $\x_j^k$ within the group $\X_j$. It is denoted by $\p^k$ for the $k$th seismogram, highlighting its waveform-specific nature. 
\end{itemize}
\begin{figure}
  \centering
  \includegraphics[width=\linewidth]{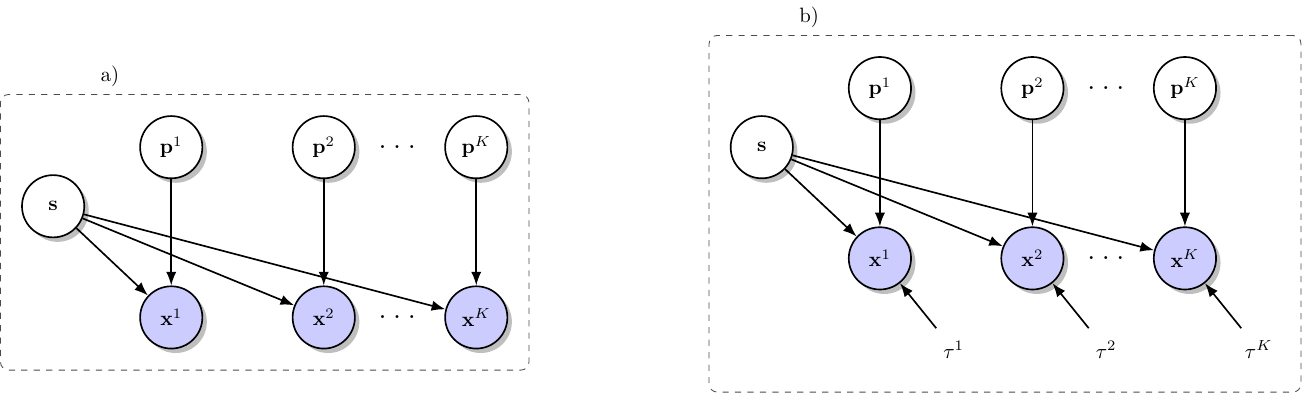}
  \caption{Generative models. (a) SymAE: this model assumes each waveform (denoted by $\x^k$) is generated from a common coherent component (represented by $\s$), along with waveform-specific nuisance components (represented by $\p^k$) that account for variations between waveforms. (b) SymAE with time-shift transformer: this enhanced model builds upon the regular SymAE by introducing a deterministic time-shift variable (denoted by $\tau^k$). This variable allows for additional flexibility in capturing coherent information from waveforms affected by temporal shifts.
  \label{fig:symae_graph}}
\end{figure}
Fig.~\ref{fig:symae_graph}a illustrates the generative model assumed by SymAE, from which the following arguments are inferred.
The source information and the path scattering information for distinct seismograms are assumed to be independent within the prior distribution. Consequently, the joint probability distribution of $\s$, $\p^k$ and $\p^l$ is expressed as the product of their marginal probability distributions
\begin{equation}
P(\s, \p^k, \p^l) = P(\s) P(\p^k) P(\p^l),
\end{equation}
for any two path codes of seismograms with indices $k$ and $l$.
In order to factor the likelihood $P(\x_i\mid\z)$, we assume the independence of seismograms conditioned on the shared 
source
information. 
For any two seismograms, indexed by $k$ and $l$, of an earthquake,  
this independence can be expressed as follows:
\begin{equation}
P(\x_j^k, \x_j^l \mid \s, \p^k, \p^l) = P(\x_j^k \mid \s, \p^k) P(\x_j^l \mid \s, \p^l).
\end{equation}
Intuitively, this indicates that, upon knowing the shared source information, the wavefield at one receiver does not inform the wavefield at another receiver. Consequently, for each source with index $j$, the likelihood can be ultimately factored as:
\begin{equation}
\label{eqn:likelihood_symae}
P(\X_j\mid\z, \phi) = \prod_{k}^{}  {P(\x_j^k\mid\s, \p^k, \phi)}.
\end{equation}

SymAE's encoder generates the approximate posterior distribution $Q(\z\mid \X_j, \theta)$ of the latent variables given the input seismograms of the earthquake. As discussed previously, this is crucial because the true posterior can be mathematically intractable. Disentanglement in the latent space allows us to factorize this posterior distribution.
To achieve factorization,
we assume conditional independence during inference of latent variables. This means that 
\begin{itemize}
\item the path effects are inferred independently using their respective seismograms
\begin{equation} Q(\p^k, \p^l \mid \X_j, \theta) = Q(\p^k \mid \X_j, \theta) Q(\p^l \mid \X_j, \theta) = Q(\p^k \mid \x_j^k, \theta) Q(\p^l \mid \x_j^l, \theta);
\end{equation}
\item and when we consider a specific earthquake, the inference of source information $\s$ is independent from 
that of the path information $\p^k$ at any receiver
\begin{equation} 
\label{eqn:spk_symae}
Q(\s, \p^k \mid \X_j, \theta) = Q(\s \mid \X_j, \theta) Q(\p^k \mid \X_j, \theta) = Q(\s \mid \X_j, \theta) Q(\p^k \mid \x_j^k, \theta). \end{equation}
\end{itemize}
By consolidating these assumptions, for each earthquake $j$, the approximate 
posterior distribution is given by
\begin{equation}
\label{eqn:posterior_symae}
Q(\z\mid\X_j, \theta) 
= Q(\s\mid\X_j, \theta)\prod_{k}^{} Q(\p^k\mid\x_j^k, \theta).
\end{equation}
In the above equation, the term $Q(\s \mid \X_j, \theta)$ represents
the coherent source information $\s$ inferred by the accumulation of information across all seismograms, and $Q(\p^k\mid\x_j^k, \theta)$ is the path (nuisance) information derived from each seismogram.
Finally, by substituting $\x_i$ with $\X_j$ in eq.~\ref{eqn:ELBO2},
and by using the expressions for likelihood and posterior from equations ~\ref{eqn:likelihood_symae} and ~\ref{eqn:posterior_symae}, we derive the ELBO for SymAE which is optimized during the training:
\begin{eqnarray}
\label{eqn:SYMAE_ELBO1}
\mathbb{L}(\theta, \phi) &=& \sum_j \int Q(\z\mid\X_j, \theta) \log\left[P(\X_j\mid\z,\phi)\right] \mathrm{d}\z - \sum_j \KL{Q(\z\mid\X_j, \theta)}{P(\z)} \nonumber \\
&   =&\sum_j\sum_k \int Q(\s\mid\X_j, \theta) Q(\p^k\mid\x_j^k, \theta) \log\left[P(\x_j^k\mid\s, \p^k, \phi)\right] \mathrm{d}\s\,\mathrm{d}\p^k 
\nonumber \\
&& 
- \sum_j \KL{Q(\s \mid\X_j, \theta)}{P(\s)} \nonumber \\
&&
- \sum_j \sum_k \KL{Q(\p^k\mid\x^k_j, \theta)}{P(\p^k)}.
\end{eqnarray}
In the above equation, the first term represents the reconstruction loss. During training, a Monte Carlo estimate is used for this term, and the next two terms are regularization terms.

\begin{figure}
  \centering
  \includegraphics[width=\linewidth]{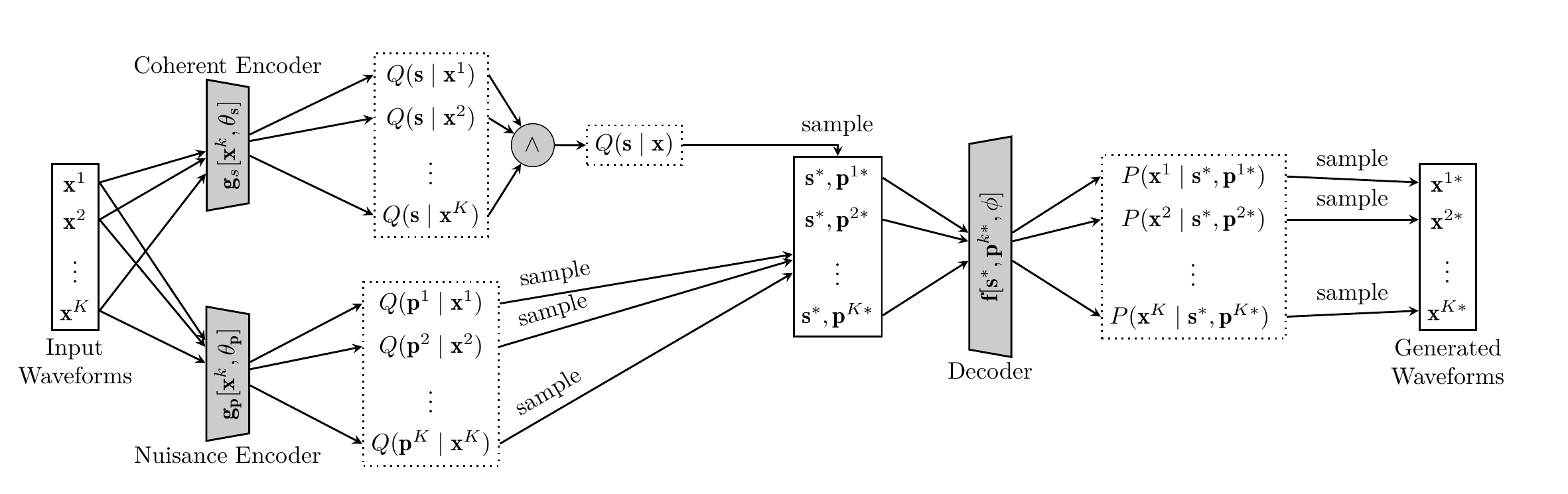}
  \caption{Disentangled representation learning for seismic waveforms using SymAE.
  \label{fig:vsymae}}
\end{figure}
\subsection{Nuisance Encoder}
The factored approximate posterior distribution $Q(\z\mid\X_j, \theta)$ in eq.\ref{eqn:posterior_symae} allows us to employ separate encoder functions for each of the decomposed terms. These separate encoders operate on the input data $\X_j$
to infer the individual latent variables that represent the source component ($\s$) and the path information components ($\p^k$) as depicted in Fig.~\ref{fig:vsymae}.
A separate convolutional network, denoted by $\g_{\p}$ with parameters $\theta_{\p}$, is employed to capture the path parameters specific to each seismogram. These parameters, denoted by $\p^k$ for $k$th seismogram, account for variations or noise present in individual seismograms that are not part of the underlying coherent information.
The network $\g_{\p}$ acts as a deterministic transformation, taking the seismogram $\x_j^k$ and its own parameters $\theta_{\p}$ as input. The output of this transformation defines the approximate posterior distribution $Q(\p^k\mid \x_j^k, \theta_{\p})$. This distribution represents the probability of the path parameters $\p^k$ given the specific seismogram $\x_j^k$ and the network parameters $\theta_{\p}$:
\begin{equation}
Q(\p^k \mid \x_j^k, \theta_{\p}) = \mathcal{N}(\p^k \mid \mathbf{g}_{\p}[\x_j^k, \theta_{\p}])
\end{equation}
Again, $\N$ denotes a normal distribution with mean and diagonal variance determined by the network output $\mathbf{g}_{\p}[\x_i^k, \theta_{\p}]$, where we denote the mean of this distribution using $\m_j^k$ for future use.

\subsection{Coherent Encoder: Accumulation of Coherent Information}
The concept of combining information from multiple sources is crucial in various scientific fields.
In \cite{tarantola2005inverse} (example 1.12),
an inverse problem is presented where independent measurements provide posterior probability distributions for a parameter. 
The combination of these independent probability distributions is achieved by first multiplying the individual distributions. Then, the resulting probability distribution is normalized to ensure that it sums up to one, representing the combined knowledge about the parameter after considering all measurements. This process, termed \emph{conjunction} (denoted using $\land$) of information states, leverages the independence of the measurements, thereby accumulating the information provided by each distribution.
It's important to remember that the process of conjunction used to combine the information from individual distributions is commutative (symmetry with respect to ordering).

Similarly, in SymAE, $Q(\s | \X_j, \theta)$ in eq.~\ref{eqn:posterior_symae} represents the accumulated coherent information about the source parameter $\s$ inferred from all seismograms in $\X_j$.
Each seismogram is considered an independent observation that is informative about $\s$. The concept of conjunction is then used to accumulate this information across all seismograms:
\begin{equation}
Q(\s \mid \X_j, \theta) = Q(\s \mid \x_j^1, \theta)\land Q(\s \mid \x_j^2, \theta)\land\cdots \propto \prod_{k}^{} Q(\s \mid \x_j^k, \theta).
\end{equation}
We model each $Q(\s \mid \x_j^k, \theta)$ as a Gaussian distribution with mean $\mc_j^k$
  and variance $\varc_j^k$.
These are determined by the output of the network $\g_s$ with parameters $\theta_s$, which takes the input seismogram $\x_j^k$ as input:
$Q(\s \mid \x_j^k, \theta_s) = \mathcal{N}(\s\mid \g_s[\x_j^k, \theta_s])$.
Given that each individual posterior distribution $Q(\s \mid \x_j^k, \theta_s)$ is Gaussian with mean $\mc_j^k$
  and variance $\varc_j^k$, it can be shown (see \cite{mlvae}) that the accumulated distribution is also Gaussian, with mean ($\mc_j$) and variance ($\varc_j$) are derived using the following equations:
\begin{eqnarray}
\varc_j^{-1} &=& \sum_{k}^{} \left(\varc_j^k\right)^{-1}, \\
\mc_j^{\text{T}} \varc_j^{-1} &=& \sum_{k}^{} \left(\mc_j^k\right)^{\text{T}} \left(\varc_j^k\right)^{-1}.
\end{eqnarray}
These equations combine the information from all individual distributions, effectively accumulating the knowledge about the source parameter $\s$ across all seismograms.
 A crucial aspect of these equations lies in using the inverse (denoted by $^{-1}$) of variance, which represents the precision. Precision indicates how informative a distribution is. A lower variance (higher precision) signifies a tighter distribution with more concentrated information around the mean. In contrast, a higher variance (lower precision) represents a broader and less informative distribution.
 The equations do not directly sum the variances because variance reflects the spread of the distribution, not its informativeness. Instead, they sum the inverses of the individual variances. This makes sense because precision tells us how much information each seismogram contributes. By adding these precisions, we effectively accumulate the total informative content from all observations.
 The second equation follows a similar logic for the mean. It calculates the product of the mean vector transposed $\mc_j^T$ with the inverse of the corresponding variance $\varc_j^{-1}$. This essentially weights the contribution of each mean based on the precision of its corresponding distribution. 
 The sum of all seismograms then combines this weighted information.
 Seismograms with higher precision (more informative) have a greater influence on the final accumulated mean $\mc_j$. 
 This formulation allows the network to implicitly down-weight waveform features with high noise (high variance, low precision). These noisy features have a weaker influence on the accumulated mean $\mc_j$. This mechanism generalizes non-linear stacking approaches where noisy samples receive lower weights. In this framework, the network effectively learns to discriminate between noisy samples and waveforms with good Signal-to-Noise Ratio (SNR), where \emph{signal} refers to coherent information, alleviating the need for manual pre-processing steps.

\subsection{Identifying Informative Seismograms}
\begin{figure}%
    \centering
    \subfloat[Seismograms with coherent source information.]{\includegraphics[width=0.45\linewidth]{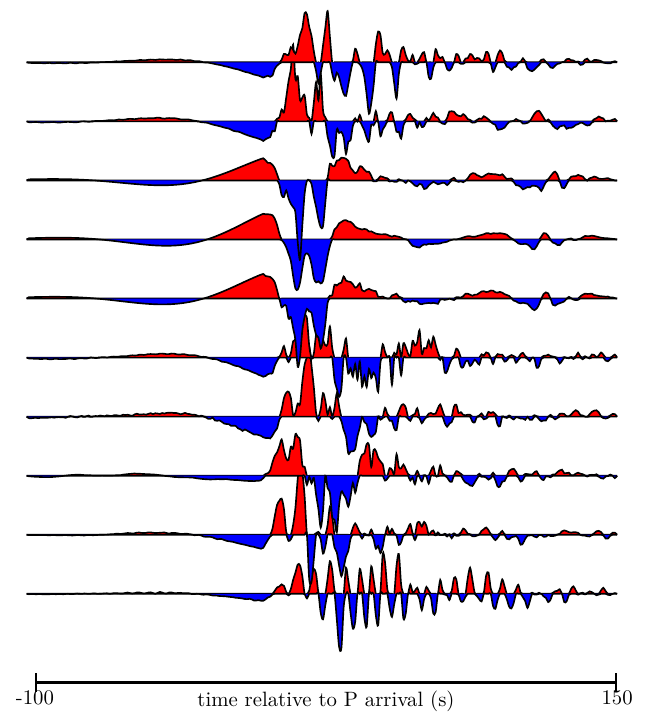}}%
    \subfloat[Seismograms lacking coherent source information.]{\includegraphics[width=0.45\linewidth]{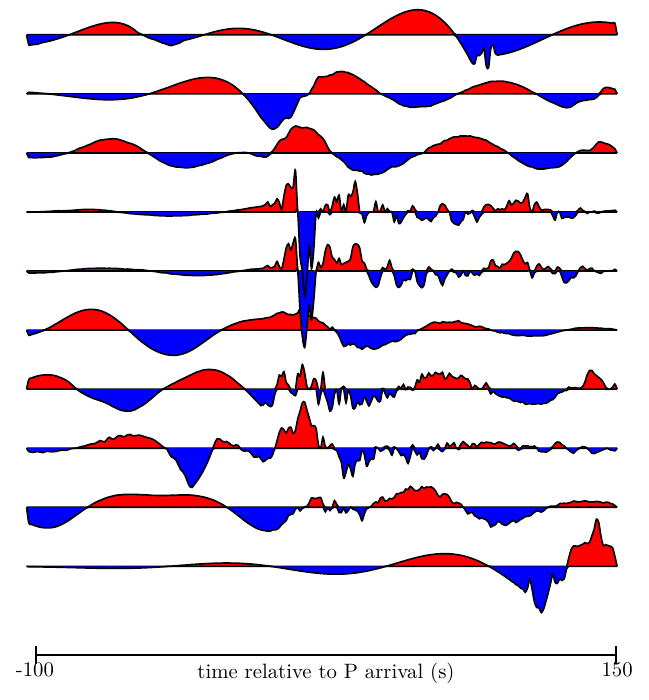}}%
    \caption{SymAE, applied here to the wavefield data from the 2013-05-24 Mw 8.3 Sea of Okhotsk earthquake, identifies seismograms that are more informative about the source by analyzing their contribution to the accumulated coherent information (see eq.~\ref{eqn:H}). Notice that seismograms with a stronger contribution and lower influence from nuisance variations are considered more informative.
}%
    \label{fig:best_seismograms}%
\end{figure}
Our approach utilizes $Q(\s \mid \x_j^k)$ to represent the knowledge about the coherent information obtained solely from the $k$-th sesimogram. In contrast, $Q(\s \mid \X_j)$ captures the accumulated coherent information derived from all seismograms of the earthquake.
After training SymAE, we can identify which specific seismograms contribute most significantly to the accumulated source information. This is accomplished by calculating the Kullback-Leibler (KL) divergence between $Q(\s \mid \X_j)$, which represents the accumulated distribution, and $Q(\s \mid \x_j^k)$, the distribution for an individual seismogram $\x_j^k$:
\begin{equation}
    \KLk[\x^k_j] = \KL{Q(\coh \mid \X_j)}{Q(\coh \mid \x_j^k)}.
    \label{eqn:H}
\end{equation}
Since KL divergence measures the distance between two distributions, seismograms that contribute more significantly to the accumulated information will have lower $\KLk$ values. In simpler terms, waveforms with lower $\KLk$ values are less affected by irrelevant path or nuisance variations and hence contribute more to the overall understanding represented by $Q(\s \mid \X_j)$.
For instance, Fig.~\ref{fig:best_seismograms}a displays seismograms that provide valuable source information of a deep earthquake, whereas Fig.~\ref{fig:best_seismograms}b shows seismograms that lack earthquake source content; it is evident that the uninformative seismograms exhibit noise, meaning they are influenced by irrelevant variations. We achieved these results by training the SymAE on a collection of high-magnitude earthquake P-wave teleseismic displacement waveforms of earthquakes listed in Tab.~\ref{tab:eq_details}. The specific details of this dataset and training process are presented later in the results section.

\subsection{Redatuming and Conditonal Generation of Virtual Seismograms}
A key advantage of the latent space disentanglement achieved in SymAE is the ability to perform conditional generation of seismograms.
In conditional generation, we can control the generation process by fixing the latent variables associated with the source information. By doing so, we can generate a collection of virtual seismograms (ensemble) that share the same underlying source characteristics while varying in their path effects.
We denote virtual or synthesized seismograms
$\y_j[\m^k_i]$, which is the mean of the Gaussian distribution, generated by the decoder $\f$ of the trained SymAE model, given by
$\f[\mc_j, \m^k_i, \phi]$.
Here, $\mc_j$ represents the mean source information for the $j$th earthquake, obtained using the coherent encoder $\g_{\s}$ that yields $Q(\s \mid \X_j)$ using seismograms $\X_j$.
 The mean path information from $\x^k_i$, the $k$th seismogram of the $i$th earthquake, is denoted using $\m_i^k$. It is derived from the encoder $\g_{\p}$ which produces $Q(\p^k \mid \x_j^k)$ using $\x_j^k$.
In seismic data processing, a key technique called redatuming aims to virtually reposition recorded seismic wavefields to a different reference level (datum). This is often done to simplify the interpretation of features within the Earth's subsurface. In the context of SymAE, we can leverage the disentangled representation to perform a specific type of redatuming: swapping the path information between different seismograms to synthesize virtual seismograms.

\begin{figure}
  \centering
  \includegraphics[width=\linewidth]{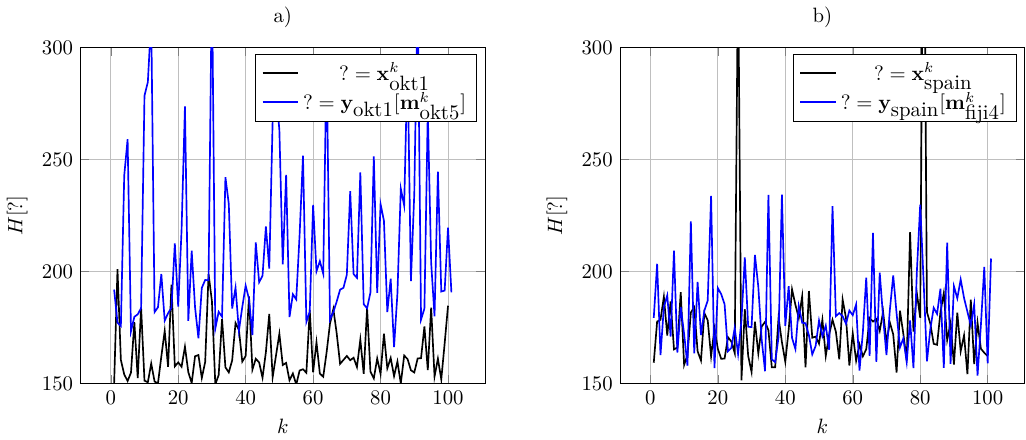}
  \caption{In the framework of source imaging, seismograms that exhibit higher $\KLk$ values (refer to eq.~\ref{eqn:H}) are considered informative. a) Virtual seismograms of the Sea of Okhotsk earthquake (Mw 8.3, referred to as okt1) in blue, generated using the path effects of another Sea of Okhotsk earthquake (Mw 6.7, referred to as okt2), are found to be less informative than the actual recorded seismograms (black). This is because okt1 and okt2 have different source mechanisms; okt2 involves a single rupture, while okt1 consists of three distinct ruptures. b) Similar to (a) but displaying the $\KLk$ values for measured and virtual seismograms of the Spain earthquake (Mw 6.3, labeled spain). In this case, both the measured and virtual seismograms (using path effects from a Mw 6.8 earthquake of Fiji region) show comparable $\KLk$ values, indicating they are equally informative. This similarity suggests that both earthquakes share characteristics.}
  \label{fig:H}
\end{figure}

\begin{figure}
  \centering
  \includegraphics[width=0.8\linewidth]{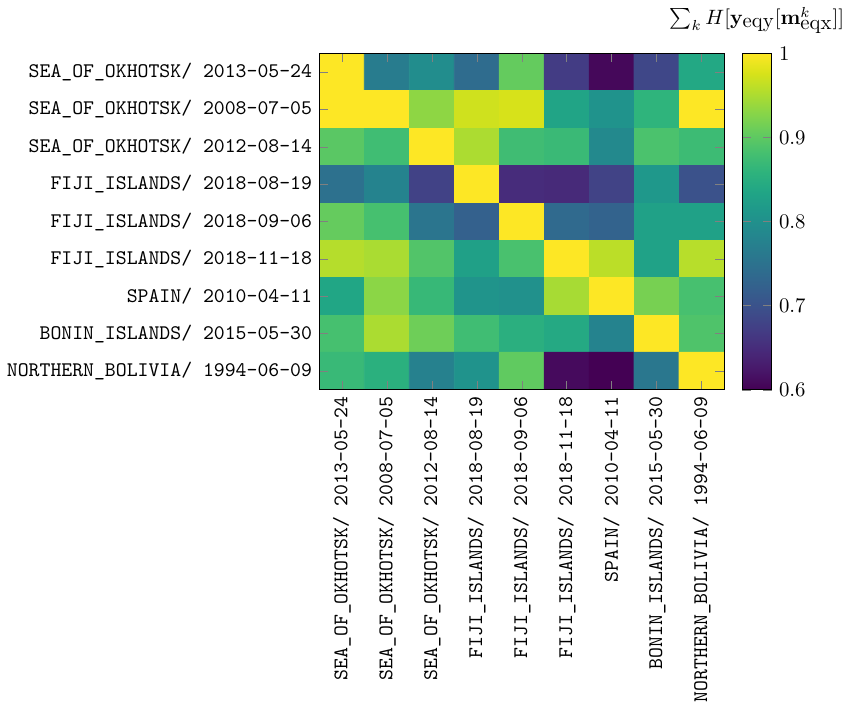}
  \caption{Heatmap illustrating the average information loss when generating virtual seismograms by combining source information from y-axis earthquakes with path information from x-axis earthquakes. Lighter colors indicate lower information loss, suggesting greater similarity between the source characteristics of the respective earthquakes.}
  \label{fig:H2}
\end{figure}

Virtual seismograms $\y_j$ generated by combining the source information of the $j$th earthquake with the path information of a different earthquake can provide insight into the role of path effects that obscure the interesting source information. For example, a clear seismogram (less affected by path effects and other noises) obtained from one earthquake could assist in producing a clear seismogram for another earthquake, potentially of smaller magnitude.
However, it is crucial to assess their overall informativeness. 
To evaluate the informativeness of a virtual seismogram, we use the eq.~\ref{eqn:H} metric based on the KL divergence:
\begin{equation}
    \KLk[\y_j[\m_i^k]] = \KL{Q(\coh \mid \X_j)}{Q(\coh \mid \y_j[\m_i^k])}.
    \label{eqn:H2}
\end{equation}
Here, $Q(\coh \mid \y_j[\m_i^k])$ represents the source information in the virtual seismogram, $\KLk[\mathbf{y}_j[\mathbf{m}_i^k]]$ quantifies the information loss when replacing the group of real seismograms $\X_j$ with the single virtual seismogram 
$\y_j[\m_i^k]$ in terms of the source information. 
A lower KL divergence indicates a better preservation of source information in the virtual seismogram.
We observed a correlation between the informativeness of generated virtual seismograms and the similarity of source earthquakes. Redatuming between similar earthquakes generally results in lower information loss compared to redatuming between dissimilar events. This suggests that earthquakes with significantly different rupture complexities, such as those with varying numbers or characteristics of seismic subevents, may exhibit greater information loss when their path effects are swapped.
Fig.~\ref{fig:H} illustrates this phenomenon using two deep earthquakes as examples.
Here, virtual seismograms generated by swapping path effects between two earthquakes even within the same tectonic setting (e.g., the Sea of Okhotsk),
as in Fig.~\ref{fig:H}a, exhibited higher information loss. 
In the case of Fig.~\ref{fig:H}b with spain and fiji4 earthquakes in Tab.~\ref{tab:eq_details}, the information loss was minimal, suggesting a resemblance between earthquakes, even if they originated from distinct tectonic settings.

To assess information loss across multiple earthquake pairs, we averaged the KL divergence $\KLk[\y_j[\m_i^k]]$ over all seismograms for each pair and visualized the results in a heatmap Fig.~\ref{fig:H2}, where higher values
after normalization indicate greater similarity between the source characteristics of the respective earthquakes.
For instance, earthquakes fiji1 and fiji2, known for their complex rupture processes with multiple subevents~\citep{JIA2020115997}, exhibited higher information loss when paired with other deep earthquakes in Tab.~\ref{tab:eq_details}.
In contrast, the earthquakes in Spain and Fiji4 showed a lower loss of information, suggesting potential similarities in their source characteristics.
Similarly, our analysis revealed similarities between earthquakes okt1 and bon1, while demonstrating a clear dissimilarity between bon1 and spain. 
We will further validate these observations by analyzing the extracted coherent source information in the results section.

\subsection{Optimal Virtual Seismograms}
We have shown that SymAE not only has the ability to generate virtual seismograms, but it can also effectively identify informative seismograms (eqs.~\ref{eqn:H} and \ref{eqn:H2}). 
Taking advantage of this dual functionality, we aim to generate virtual seismograms that are enriched with task-relevant source information while simultaneously minimizing the influence of path factors. Essentially, we used a trained SymAE model to create a virtual seismogram that best represents the underlying source characteristics, effectively filtering out noise and irrelevant information that may be present in real-world data.
The generation of this virtual seismogram translates into an optimization problem, where we search the space of possible nuisance codes (denoted by $\m$) to achieve this goal. This optimization problem can be formulated as follows:
\begin{eqnarray}
{\m}_j &=& \operatorname*{arg\,min}_{\m} \; \KL{Q\left(\s \mid \X_j\right)}{Q(\s \mid \y_j[\m])} + \alpha \|\y_j[\m]\|_{1},\nonumber \\
&& \text{subject to}\quad\sum_{n=1}^{D}(\y_j[\m])_n=0\quad\text{and}\quad\frac{1}{D}\sum_{n=1}^{D}{\left((\y_j[\m])_n-\frac{1}{D}\sum_n (\y_j[\m])_n\right)^2}=1.
\end{eqnarray}
 The solution we seek is the optimal nuisance code, denoted by $\m$, corresponding to the $j$th earthquake. By manipulating this code within the optimization framework, we influence the characteristics of the generated virtual seismogram $\y_j[\m]$, which is the mean of decoder output $\f[\mc_j, \m]$.
A key term in the objective function is the KL divergence, denoted by 
$\KL{Q(\s \mid \X_j)}{Q(\s \mid \y_j[\m])}$. This term measures the difference between the information about the coherent source (represented by $\s$) extracted from all measured seismograms
of the earthquake $Q(\s \mid \X_j)$ and
the information obtained from the generated virtual seismogram $Q(\s \mid \y_j[\m])$. 
Minimizing this divergence ensures that the virtual seismogram retains the coherent information of the earthquake.
Another crucial term promotes sparsity in the virtual seismogram. This term
involves the $\ell^1$-norm of the predicted virtual seismogram. A higher $\ell^1$-norm penalty encourages a virtual seismogram with fewer non-zero values, potentially leading to a simpler representation with less influence from nuisance variations. The parameter $\alpha$ controls the weight given to this sparsity term, allowing us to adjust the trade-off between minimizing KL divergence and achieving a sparse virtual seismogram.
%
%
Other relevant constraints, denoted by ``subject to'' in the formulation, include
enforcing a zero mean for the virtual seismogram. This translates to requiring the sum of all elements in the D-dimensional vector $\y=(y_1,\ldots, y_D)^{\mathrm{T}}$ to equal zero.
The second constraint ensures unit variance for the virtual seismogram. 
These constraints align the generated virtual seismogram with real seismograms, which exhibit zero mean and unit variance characteristics after preprocessing.
In the results section, we show that generating a virtual seismogram for each earthquake effectively preserves important source information while reducing the impact of noise and other irrelevant variations.

\section{Symmetric Autoencoder with Time-shift Transformer}
\begin{figure}
  \centering
  \includegraphics[width=\linewidth]{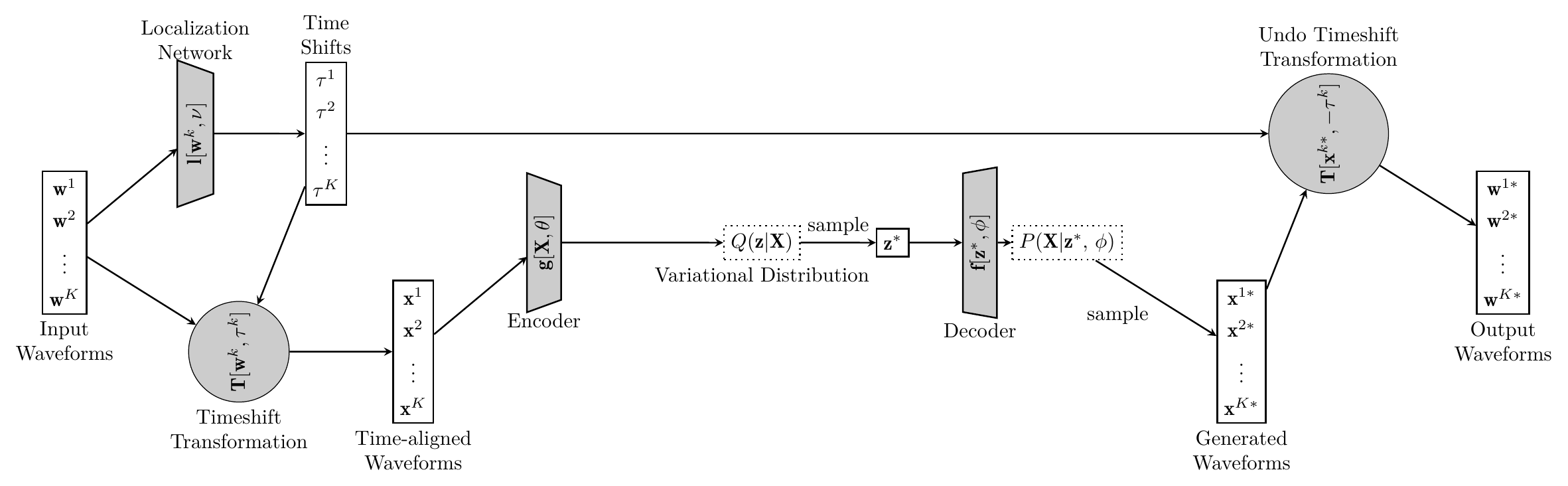}
  \caption{Architecture of the SymAE with time-shift transformer. The model incorporates a localization network to estimate optimal time shifts for input waveforms, followed by the core SymAE components for latent-space representation and reconstruction. An inverse time-shift operation is applied to the reconstructed waveforms to align them with the original waveforms.
  \label{fig:vsymae-timeshift}}
\end{figure}

Seismograms often exhibit variations in arrival times due to factors such as propagation path differences. This can hinder the effective application of the SymAE architecture discussed in the previous section, which has difficulty compressing time-shifted waveforms into a low-dimensional latent space. To address this challenge, we introduce a timeshift transformer module that aligns waveforms before encoding. We show that by learning optimal time shifts, the model effectively extracts underlying coherent information and generates accurate reconstructions.

The architecture of SymAE with time-shift transformer is illustrated in Fig.~\ref{fig:vsymae-timeshift}, while its corresponding generative model is shown in Fig.~\ref{fig:symae_graph}b, where time shifts are treated as deterministic variables.
To account for potential time shifts in input waveforms, we introduce the notation $\W_j = \{\w_j^k\}$ to represent time-shifted seismograms. A localization network $\mathbf{l}$ with parameters $\nu$ initially processes these waveforms to determine optimal time shifts $\tau_j^k=\mathbf{l}[\w_j^k, \nu]$, which are then applied to generate time-aligned versions $\X_j = \{\x_j^k\} = \{\mathbf{T}[\w_j^k, \tau_j^k]\}$. Here, $\mathbf{T}$ denotes the time-shift operation, which we implement in the frequency domain using the Fourier-shift property.
 The core SymAE model subsequently operates on these aligned waveforms to learn a disentangled latent space representation. 
 The loss function, adapted from the standard VAE formulation in Eq.~\ref{eqn:ELBO2}, incorporates a reconstruction term, a KL divergence regularizer, and a time-shift regularization term: 
\begin{eqnarray}
\label{eqn:ELBO_timeshift}
\mathbb{L}(\theta, \phi, \nu) &=& \sum_j \int Q(\z\mid\W_j, \theta) \log\left[P(\W_j\mid\z,\phi)\right] \mathrm{d}\z - \sum_j \KL{Q(\z\mid\W_j, \theta)}{P(\z)} \nonumber \\
&& + \gamma \sum_j \sum_k (\tau_j^k)^2,\quad\text{where}\quad \tau_j^k=\mathbf{l}[\w_j^k, \nu]\quad \text{and}\quad \x_j^k = \mathbf{T}[\w_j^k, \tau_j^k]
\end{eqnarray}
An inverse time-shift operation using $-\tau_j^k$ is applied to the decoder-produced waveforms $\f$. The mean of the likelihood $P(\W_j\mid\z,\phi)$ results from this inverse time-shift operation on the decoder output $\f[\z, \phi]$. Where as, the posterior $Q(\z\mid\W_j, \theta)$ is obtained by applying the encoder after time-shifting each input waveform using the corresponding $\tau_j^k$.
The time-shift regularization term penalizes excessive time shifts, ensuring stability and interpretability of the model.
The loss function is minimized through backpropagation to optimize the model parameters $\phi$, $\theta$ and $\nu$.

\begin{figure}
  \centering
  \includegraphics[width=\linewidth]{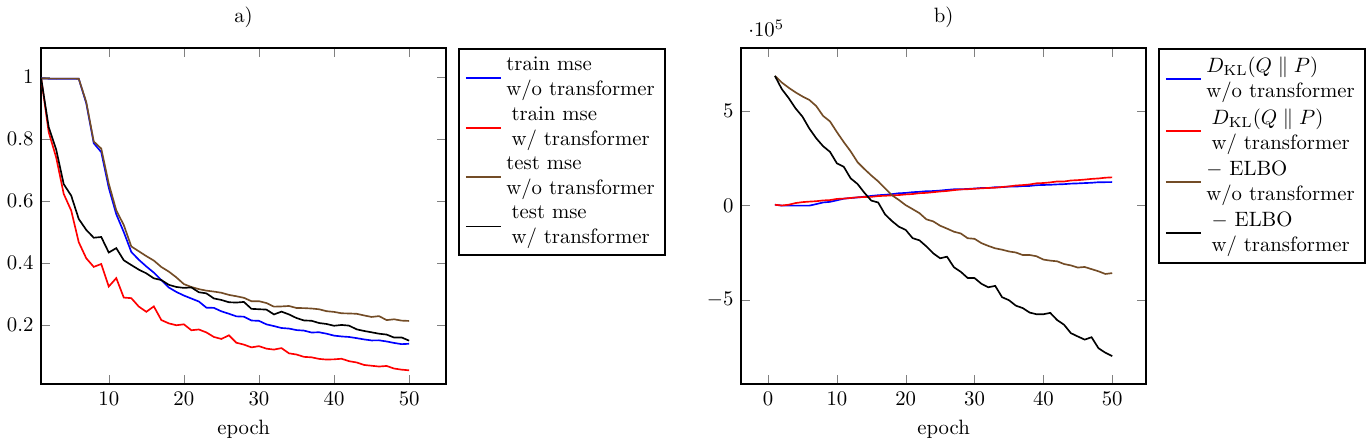}
  \caption{ELBO and KL divergence $\KL{Q}{P}$ during SymAE training with and without time-shift transformer using noiseless seismograms (Figs.~\ref{fig:syn_exp1}e and \ref{fig:syn_exp1}f). The time-shift transformer SymAE model exhibits lower reconstruction loss (mse).
  }
  \label{fig:loss1}
\end{figure}

\begin{figure}%
    \centering
    \subfloat[With noiseless seismograms (Figs.~\ref{fig:syn_exp1}e and \ref{fig:syn_exp1}f) without using time-shift transformer]{\includegraphics[width=\linewidth]{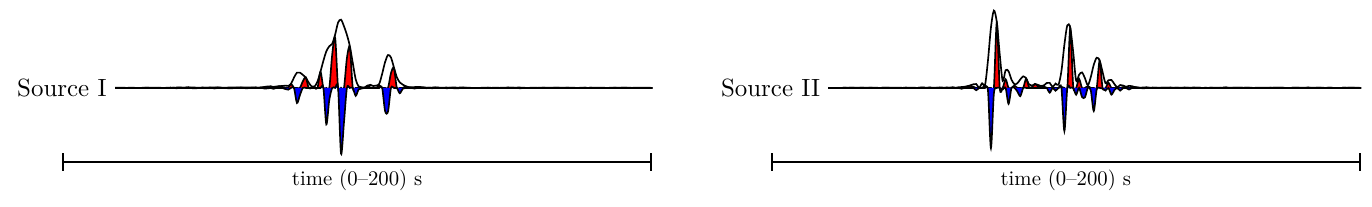}}%
  
    \subfloat[With noisy seismograms (Figs.~\ref{fig:syn_exp1}g and \ref{fig:syn_exp1}h) without using time-shift transformer]{\includegraphics[width=\linewidth]{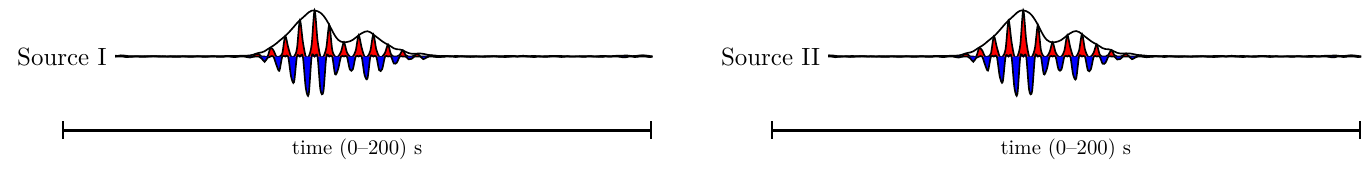}}%
  
    \subfloat[With noiseless seismograms (Figs.~\ref{fig:syn_exp1}e and \ref{fig:syn_exp1}f) using time-shift transformer]{\includegraphics[width=\linewidth]{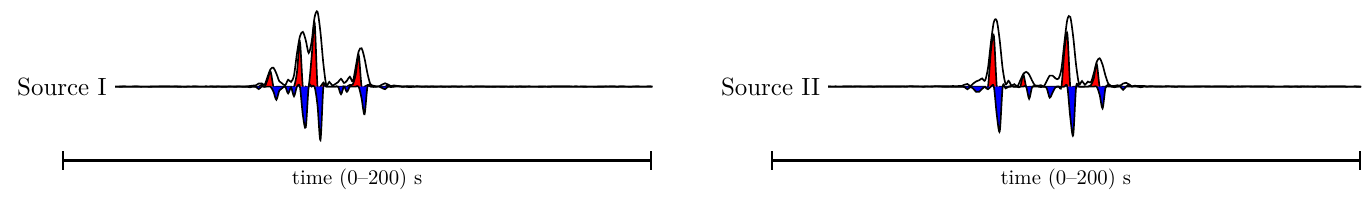}}%

    \subfloat[With noisy seismograms (Figs.~\ref{fig:syn_exp1}g and \ref{fig:syn_exp1}h) using time-shift transformer]{\includegraphics[width=\linewidth]{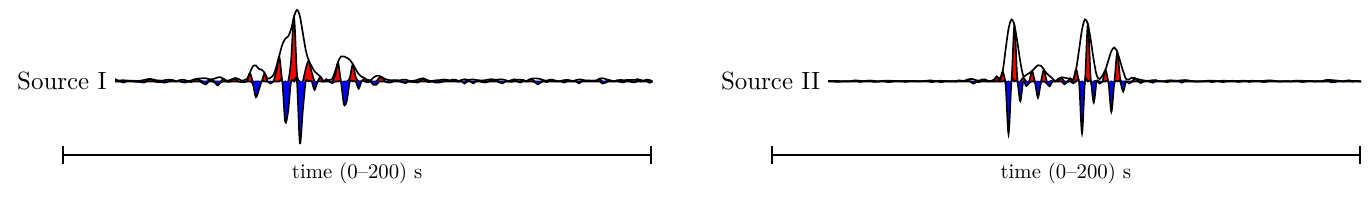}}%
    \caption{SymAE extracted coherent source information through latent space optimization for the synthetic experiment. True sources I and II are plotted in Fig.~\ref{fig:syn_exp1}a and Fig.~\ref{fig:syn_exp1}b, respectively.}%
    \label{fig:syn_result}%
\end{figure}

\begin{figure}
  \centering
  \includegraphics[width=\linewidth]{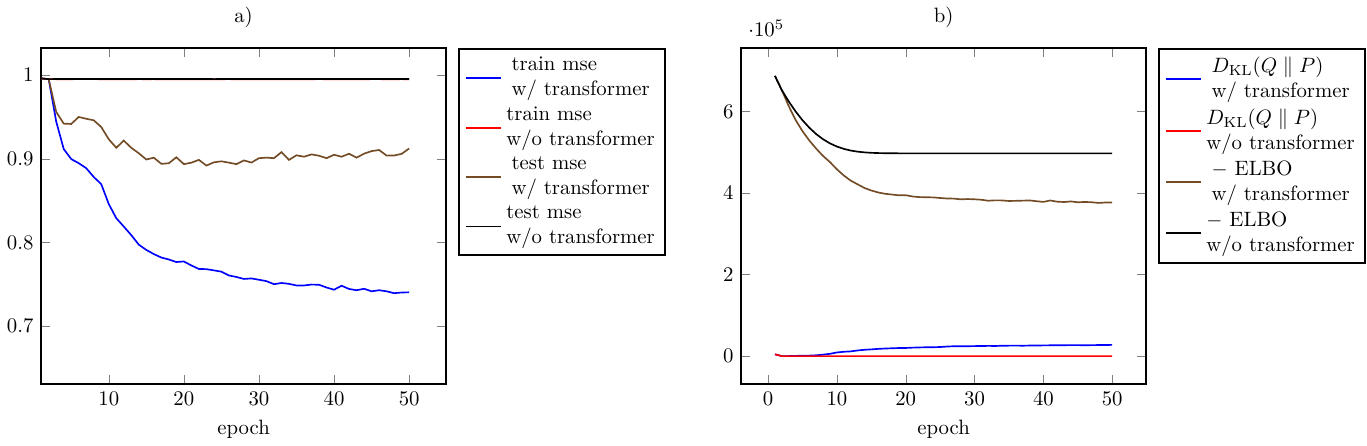}
  \caption{Same as Fig.~\ref{fig:loss1}, but using noisy seismograms in Figs.~\ref{fig:syn_exp1}g and \ref{fig:syn_exp1}h.}
  \label{fig:loss2}
\end{figure}

\section{Results}

To evaluate the proposed SymAE model, synthetic seismograms incorporating noise and time shifts were generated (Fig.~\ref{fig:syn_exp1}). The SymAE model was trained using the ADAM optimizer~\citep{kingma2014adam}. Fig.~\ref{fig:loss1} illustrates the ELBO loss during training for standard and time-shift transformer SymAE models using noiseless data. The latter exhibited lower reconstruction loss, and the extracted coherent source information (Figs.~\ref{fig:syn_result}a and ~\ref{fig:syn_result}c) closely matched the ground-truth source signatures in both cases. However, in the presence of noise, the standard SymAE model performed poorly, as evidenced by the ELBO plot in Fig.~\ref{fig:loss2}. In contrast, the time-shift transformer SymAE demonstrated superior performance, effectively extracting coherent source information, as plotted in Fig.~\ref{fig:syn_result}d.

\begin{figure}
  \centering
  \includegraphics[width=\linewidth]{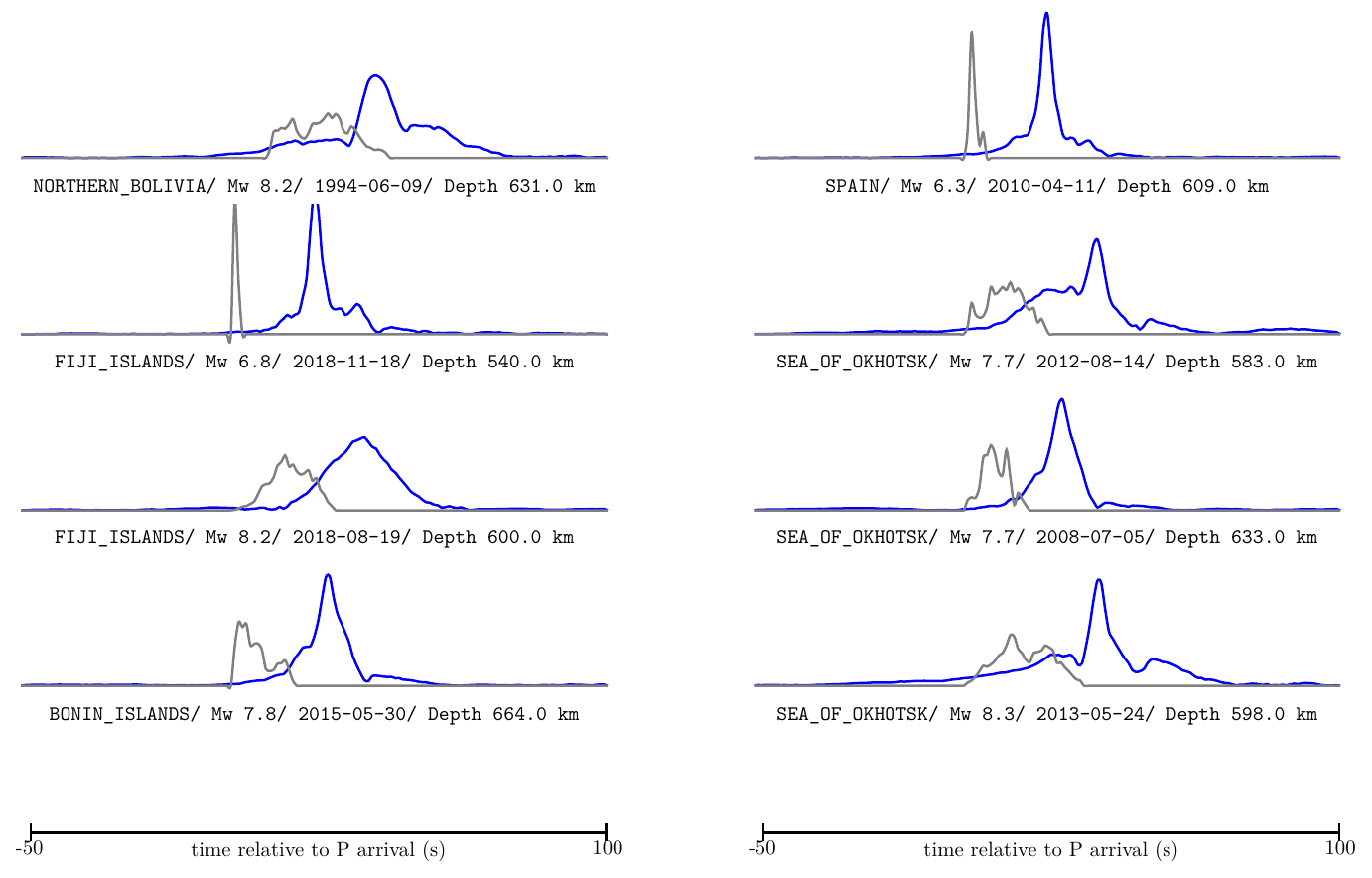}
  \caption{
Comparison of
coherent source information extracted using SymAE (blue) for complex deep earthquakes compared to traditional deconvolution techniques (gray).}
  \label{fig:deepeq}
\end{figure}

Fig.~\ref{fig:deepeq} illustrates the effectiveness of the proposed SymAE method in extracting coherent source information of deep earthquakes (Tab.~\ref{tab:eq_details}) compared to traditional deconvolution techniques~\citep{vallee2011scardec}. Here, SymAE is trained on roughly 5000 displacement seismograms (all components) for each earthquake. The SymAE-derived source time function exhibits a clearer and more detailed representation of the earthquake rupture process, capturing complex features due to multiple events that are often obscured and smoothed out in traditional deconvolution approaches. This improvement is particularly evident in the case of earthquakes like bon1, fiji4, and spain. The half-durations of the source functions extracted using SymAE closely align with those determined from the raw seismograms.
Importantly, the extracted source-function similarities align closely with the relationships previously observed in the information loss heatmap in Fig.~\ref{fig:H2}. This further validates the effectiveness of our proposed optimization approach in extracting the underlying characteristics of the earthquakes.

\section{Conclusions}
This study introduces a novel approach to extracting coherent information from seismic wavefields using a Symmetric Autoencoder (SymAE) with a time-shift transformer. By disentangling the latent space into source and path components, the proposed method effectively isolates coherent seismic signals, reducing the impact of noise and other unwanted variations. The ability to quantify waveform informativeness is a key contribution, enabling the identification of data that significantly contributes to the overall coherent signal.
Our experiments demonstrate the effectiveness of SymAE in extracting coherent source information, particularly in the presence of noise and complex seismic phenomena. The generated virtual seismograms provide valuable insights into the underlying earthquake processes. The proposed framework holds promise for a wide range of seismological applications that require the extraction of coherent signals from noisy data.

Future research could explore the application of SymAE to other seismic datasets, such as ambient noise tomography. 
Additionally, incorporating more complex physical models into the SymAE framework could lead to further improvements in the extraction of coherent information.
We emphasize the importance of evaluating generative models, such as SymAE, within the context of their intended applications. We argue that a model's performance should be judged on the basis of its ability to produce data that are physical and valuable for specific seismological tasks, as generic metrics may not capture the nuances required for these applications. \cite{Theis2016} recommended to use evaluation strategies that are tailored to the specific applications of generative models.

\section{Acknowledgments}
This work is funded by the Science and Engineering Research Board, Department of Science and Technology, India (Grant Number SRG/2021/000205).
The authors would like to express their sincere gratitude to Matt Li and Laurent Demanet for their valuable insights on variational autoencoders, and to Sanket Bajad and Tiente Rengneichuong for discussions. We also extend our thanks to Isha Lohan and Madhusudan Sharma for their contributions to earthquake data preprocessing.
Our research was significantly aided by the utilization of the Flux.jl deep learning framework, which enabled us to implement and train the SymAE model. We also acknowledge the utilization of the Julia programming language, specifically version 1.9.0, as well as Pluto notebooks, which played a pivotal role in the development of our numerical simulations and data analysis presented throughout this work.

  
  \section{Data Availability}
All data and codes used is this study are open access. 
The seismological data is primarily sourced from the Incorporated Research Institutions for Seismology Data Management Center (IRIS-DMC), where the data was  acquired from an array of institutions and seismic data centers, including AUSPASS, BGR, EMSC, ETH, GEOFON, GEONET, GFZ, ICGC, IESDMC, INGV, IPGP, IRIS, IRISPH5, ISC, KNMI, KOERI, LMU, NCEDC, NIEP, NOA, ODC, ORFEUS, RASPISHAKE, RESIF, RESIFPH5, SCEDC, TEXNET, UIB-NORSAR, USGS, and USP.

\bibliographystyle{unsrtnat}
\bibliography{example}  






\end{document}